# On the feasibility of a component-based approach to predict aerodynamic noise from high-speed train bogies

Eduardo Latorre Iglesias[1*], David Thompson[2], Jorge Muñoz Paniagua[3], Javier García García[3]

[1] Escuela Técnica Superior de Ingeniería y Sistemas de Telecomunicación, Universidad Politécnica de Madrid,
C/ Nikolas Tesla s/n (Campus Sur), 28031, Madrid, Spain

[2] Institute of Sound and Vibration Research, University of Southampton, Southampton, SO17 1BJ, UK.

[3] Escuela Técnica Superior de Ingenieros Industriales, Universidad Politécnica de Madrid,
C/ José Gutiérrez Abascal, 2, 28006, Madrid, Spain

[*]Corresponding author:
e-mail address: eduardo.latorre.iglesias@upm.es (E. Latorre Iglesias)

**Abstract.** At speeds above 300 km/h, aerodynamic noise becomes a significant source of railway noise. In a high-speed train, the bogie area is one of the most important aerodynamic noise sources. To predict aerodynamic noise, semi-empirical component-based models are attractive as they allow fast and cheap calculations compared with numerical methods. Such component-based models have been previously applied successfully to predict the aerodynamic noise from train pantographs which consist of various cylinders. In this work, the feasibility is evaluated of applying them to the aerodynamic noise from bogies, which consist of various bluff bodies. To this end, noise measurements were carried out in an anechoic wind tunnel for different flow speeds using simple shapes representing various idealised components of a bogie. These results have been used to adjust the empirical constants of the component-based prediction model to enable its application to the bogie case. In addition, the noise from a 1:10 scale simplified bogie mock-up was measured and used for comparison with the prediction model. The results show good agreement between measurements and predictions for the bogie alone. When the bogie is located between two ramps approximating a simplified bogie cavity the noise is underpredicted at low frequencies. This is attributed to the noise from the cavity which is not included in the model.

## 1 Introduction

High-speed trains operate at speeds of 300 km/h or greater in a number of countries. Above about 300 km/h, aerodynamic noise becomes a significant source of railway noise [1]. The aerodynamic noise is produced by the interaction of the incoming air flow with different components or areas of the train, in particular the bogies, pantograph, inter-coach cavities, and the train nose. In the case of the bogies, the aerodynamic noise is influenced by the complex turbulent flow recirculation around the bogie area [2]. The leading bogie of the train contributes more aerodynamic noise than the others, as it is exposed to higher incoming air flow speeds [3-6]. For the remaining bogies, although the presence of upstream bogies or equipment significantly reduces the incident flow speed, their contribution can still be significant, as there are many bogies along the train [7-9]. The estimated contribution of the aerodynamic noise from all the bogies to the overall aerodynamic noise of a high-speed train during pass-by is around 15 dB greater than the contribution from the pantograph [2]; the pantograph produces a significant noise peak during the pass-by but it has a limited duration. In terms of frequency content, the aerodynamic noise from the bogie region is broadband. Results reported by Mellet et al. [9] from microphone array measurements on French high-speed trains indicated that bogie aerodynamic noise has a significant contribution to the A-weighted sound pressure between 315 Hz and 2 kHz, and especially below 1 kHz [9]. From measurements on Japanese trains at 320 km/h, Uda et al. [10] found that bogie aerodynamic noise played a greater role than the rolling and equipment noise between 125 and 500 Hz. From scale model tests, the noise from the bogie cavity was found to be predominant for frequencies below 100 Hz (at full scale) [11, 12].

The bogie aerodynamic noise can be assessed by measurements or predictions (an extensive review of assessment methods is given in [13]). If an experimental approach is used, it can be based on field measurements or wind tunnel tests. Field tests allow the noise from a real train in operation to be measured but they have drawbacks such as the availability of the train and track, and the difficulty to separate aerodynamic noise from rolling noise in the bogie region. On the other hand, anechoic wind tunnel tests have several advantages: only the aerodynamic noise is present, so there is no need for source separation; different configurations can be quickly tested using prototypes to evaluate low-noise designs; and by testing different configurations it is possible to assess the contribution of the various bogie components to the overall noise. However, due to the limitation in the wind tunnel size, scaled models have to be used and the Reynolds numbers obtained might not be representative of those of the real case, in which case the aerodynamic phenomena will not scale correctly. Moreover, not all facilities can reach the required flow speed to represent the real conditions. In addition, these tests are expensive. Examples are found in the literature of anechoic wind tunnel tests to assess bogie aerodynamic noise [14-16], to estimate the contribution of different bogie components [12, 17] and bogie configurations [18] and to assess the noise reduction obtained when applying counter measures [19, 20]. Field tests and anechoic wind tunnel measurements can also be combined with prediction models to assess bogie aerodynamic noise. For instance, Kitagawa et al. used the prediction tool TWINS [21] to calculate the rolling noise produced by a Shinkansen (Japanese high-speed train) and wind tunnel tests with a scaled train to estimate the aerodynamic noise produced by the bogies [22, 23].

Numerical predictions using computational fluid dynamics (CFD) allow the bogie aerodynamic noise to be estimated without the need for a real train or expensive test facilities. Examples of using large eddy simulations (LES) and delayed detached-eddy simulations (DDES) together with suitable acoustic analogy for calculating the noise generated by a simplified bogie are found in [24, 25]. However, the application of numerical models to large domains with complex geometry and flow behaviour is still extremely costly. The computational cost can be significantly reduced by the application of semi-empirical component-based prediction models. In these models, a database is built from experiments or numerical simulations of the noise produced by simple geometries that are used to approximate components of the whole system (e.g. pantograph, bogie). This allows the empirical constants of the model to be adjusted. The overall noise is calculated as the incoherent sum of the noise produced by each component. Models of this kind have been applied to predict the aerodynamic noise produced by aircraft landing gears [26] and train pantographs [27, 28], which are mainly produced by vortex shedding phenomena from cylinders. Promising results have been obtained, allowing a quick evaluation of different configurations and the assessment of the contribution of each component to the overall noise. However, there are additional challenges for its application to predict aerodynamic noise from a high-speed train bogie because the noise generation phenomena for more compact bluff bodies are different and the noise generated is not tonal due to vortex shedding, but broadband [29].

This work aims to investigate the feasibility of the application to a train bogie of a semi-empirical component-based model originally developed for train pantograph aerodynamic noise. The model for the pantograph case presented in [27] is applied to the bogie case. To this end, results are presented from noise measurements carried out in an anechoic wind tunnel for flow speeds ranging between 20 and 50 m/s. During these tests, noise was measured using simple geometries to represent bogie components and the results are used to adjust the empirical constants of the model. Measurements were also carried out using a 1:10 scale simplified bogie and the results are compared with predictions from the model. In practice, the effect of the (moving) ground beneath the bogie will modify the incoming flow conditions and will have a significant impact on the radiated noise [30]. For practical reasons, however, this could not be considered in the current experimental set-up.

The paper is organized as follows: the existing component-based model based on that applied to the pantograph case is summarised in Section 2. In Section 3, details are given about the wind tunnel tests; the experimental set-up is described, and the measurement results obtained using simple shapes to approximate bogie components are presented. The empirical constants used in the prediction model are adjusted based on these noise measurements and are also presented in Section 3. Section 4 includes the set-up and results obtained during the anechoic wind tunnel tests made using a 1:10 scale simplified bogie mock-up, alone and inside a simplified cavity. The comparison between predictions and measurements is included in Section 5. Finally, the conclusions are given in Section 6.

## 2   Component-based prediction model

The semi-empirical component-based prediction model is based on the assumption that the mean square sound pressure spectrum radiated by a structure exposed to an incoming flow can be expressed as the incoherent sum of the spectra of each individual component, as follows [28]:

$$\overline{p^2}(x,f) = \frac{(\rho_0 c_0^2)^2}{16} \sum_i \frac{M_i^{\alpha_i} \eta_i S_i F_i(f)}{R_i^2} \frac{D_{rad_i}(\psi,\phi)}{(1-M\cos(\theta))^4} \quad (1)$$

where $\rho_0$ is the air density, $c_0$ is the speed of sound, $M_i = U_i/c_0$ is the Mach number (where $U_i$ is the incident flow speed), $R_i$ is the distance between component and receiver, $\alpha_i$ is the speed exponent, $\eta_i$ is the amplitude factor, $S_i$ is the total surface area of the bogie component, $F_i(f)$ is a normalized frequency spectrum obtained using empirical relations and the subscript $i$ refers to each component. The directivity function $D_{rad_i}(\psi,\phi)$ depends on the angles between the lift vector acting on the component and the position vector of the receiver with respect to the component position; $\psi$ is the angle in the $x$-$y$ plane and $\phi$ is the angle in the $y$-$z$ plane, where $x$ is the airflow direction, $y$ is the lateral direction and $z$ is the height. The convective amplification effect for a noise source depending on the observer position [31] is accounted for through the factor $(1-M\cos(\theta))^4$, where $\theta$ is the angle between the flow direction and the observer position.

The amplitude factor $\eta_i$ for each component $i$ is chosen to fit the noise level to a reference spectrum. The normalized reference spectrum is obtained by fitting the shape of a given function to that of the reference measurement. Two different functions, $F_p(f)$ and $F_b(f)$ are used to model the normalized spectrum so $F_i(f)$ is obtained as the sum of these two. If vortex shedding noise is produced by the component, a "haystack-like" spectrum is used to approximate the noise from the vortex shedding peak, given by [28]

$$F_p(f) = \frac{a_{p1}}{\left(\left(f_0/f\right)^2 - \left(f/f_0\right)^2\right)^2 + a_{p2}} \quad (2)$$

where $f_0$ is the vortex shedding frequency that will also be the frequency of the peak of the normalized function, and $a_{p1}$ is a constant used as normalization factor to force the integral of $F_p(f)$ over frequency to be unity,

$$a_{p1} = \frac{1}{\int F_p(f) df} \quad (3)$$

The constant $a_{p2}$ is used to control the bandwidth of the function and is obtained from

$$a_{p2} = \frac{1}{2.25}\left(\frac{B_f}{f_0}\right)^2 \quad (4)$$

where the relative bandwidth, defined as $B_f/f_0$, is obtained as the relative width of the function at a level of 10 dB below the peak value.

A second normalized spectrum is required to model the broadband noise. This function is also used for components producing broadband noise without a distinguishable tonal peak. The following function is chosen to provide a broader "haystack-like" spectral shape [28]:

$$F_b(f) = \frac{a_{b1}}{1 + \left(f/f_0\right)^{n_2} + \left(\frac{n_2}{n_1}\right)\left(f/f_0\right)^{-n_1}} \quad (5)$$

This function is arranged to have its maximum value at the same frequency $f_0$ as the vortex shedding, where present. The value of the constant $a_{b1}$ is chosen to ensure that the integral of $F_b(f)$ over frequency is equal to unity, as explained above for $a_{p1}$. For broadband noise it is useful to control the slope of the function separately for frequencies below and above $f_0$. This is achieved by modifying the values of the constant $n_1$ for frequencies below $f_0$ and $n_2$ for frequencies above $f_0$. The ratio $n_2/n_1$ is present in the denominator to ensure that the maximum of the function occurs at the frequency $f_0$ when $n_1 \neq n_2$.

## 3     Simple shapes: experiments and model adjustment

A series of measurements has been carried out in the anechoic wind tunnel of the Institute of Sound and Vibration Research (ISVR) in the UK [32]. A smooth air flow with low turbulence is delivered through a rectangular nozzle with dimensions 0.5 × 0.35 m. The test samples consisted of various simple shapes, representing simplified bogie components, as well as a simplified 1:10 scale bogie. Figure 1 shows the test set-up, which is common for all the test cases. The test samples were attached to a stiff plywood panel surrounded by a baffle that was used to minimize the noise generated by flow interaction with the edges of the panel. The baffle was aligned flush with the edge of the nozzle to ensure smooth flow delivery and avoid flow separation. An end view of the set-up is shown in Figure 1(a) and a side view is shown in Figure 1(b). The nozzle is located on the left side (Figures 1(b) and 1(c)) with air flowing from left to right. In the configuration shown in Figure 1(a) a bogie model is attached to the rear of the baffle (see Section 4). For the tests with the simple shapes, the baffle is relocated to align with the far side of the nozzle (right-hand side in Figure 1(a)) and the shapes are attached on the other side of the baffle, as indicated in Figure 1(b). A microphone (circled in Figure 1(a)) was attached to the arc-shaped metal structure above the baffle in a direction normal to the flow opposite the centre of the test sample. Figure 1(c) shows the distance from the centre of the test samples to the nozzle outlet (0.7 m) and to the microphone (1.4 m). This microphone is used for model adjustment and comparison with predictions (as this is placed on the radiation axis the convective amplification and shear layer refraction effects can be neglected).

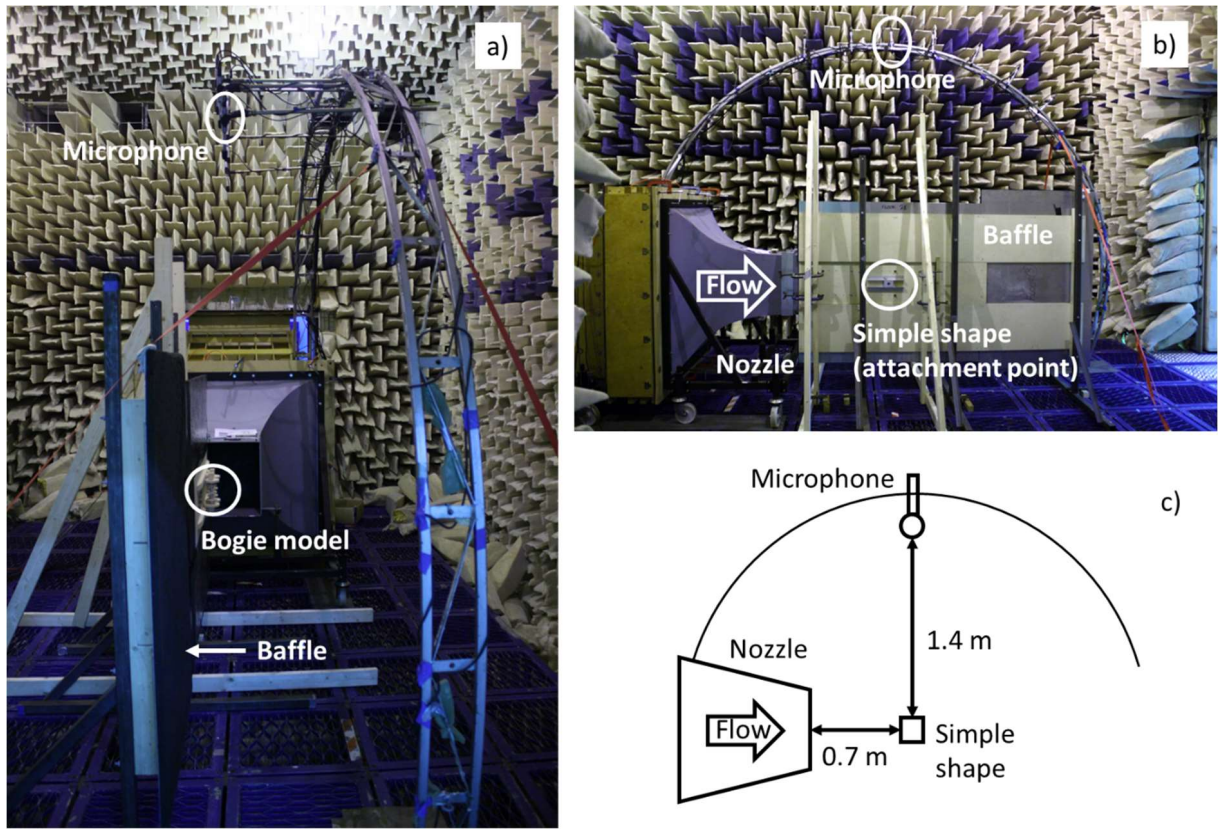

**Fig. 1.** Experimental set-up. (a) End view; (b) side view; (c) sketch showing the relative position between the centre of the sample, the nozzle and the microphones.

The measured noise was recorded over 10 s at a sampling frequency of 48 kHz. To obtain the narrow band frequency spectra, a Hanning window of 8192 samples was used with overlap of 50%, yielding a frequency resolution $\Delta f$ = 5.85 Hz.

### 3.1 Anechoic wind tunnel noise tests

Measurements were performed first on various simple shapes made of solid plywood, used to approximate bogie components. A disc was used to represent wheels and disc brakes, and two rectangular cuboids with different dimensions, one to represent the motor and the other, with rounded corners, to represent the gearbox; all of them had dimensions of approximately 1/5 of the size of real bogie components. A cube with sharp edges was also used. These objects are shown in Figure 2 together with their main dimensions. Holes were introduced as shown to allow a strut to be inserted to support the objects. The characteristic dimension $D$ is defined in each case as that perpendicular to the flow direction and parallel to the radiation axis. These idealised geometrical shapes are used to simplify the experiments while keeping the main aerodynamic features of the corresponding objects in a bogie. It is expected that the lack of small details will lead to underestimates of the high-frequency noise in comparison with that produced by the real bogie components. The noise spectra measured with these simple shapes is used to adjust the value of the empirical constants in the prediction model.

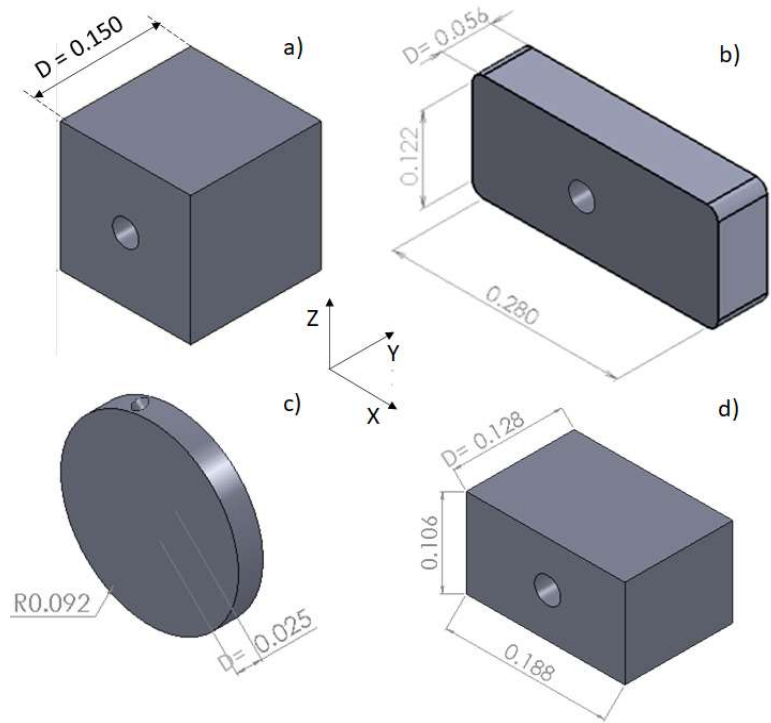

**Fig. 2.** Simple shapes used to approximate the main bogie components. Dimensions are given in mm. (a) Cube; (b) long rectangular cuboid representing a gearbox; (c) disc representing a wheel or disc brake; (d) rectangular cuboid representing a motor. In the centre, coordinates used during the experiments. The flow direction corresponds to the *x*-axis in each case.

### 3.1.1 Experimental set-up

Figure 3 shows the front view of the specific experimental set-up used for the simple shapes. These were supported using a long horizontal strut with diameter 12 mm attached to the baffle at one end and supported by a stand outside the flow at the other end. The strut was wrapped with porous foam of thickness 6 mm to avoid vortex shedding and minimize the background noise. To ensure clean incident flow, the different shapes were installed with a 50 mm gap between them and the baffle. This ensures that their location is not inside the turbulent boundary layer developed along the baffle. The nozzle provided flow with different flow speeds (20, 25, 31.5, 40 and 50 m/s) covering a Reynolds number range $3.42 \times 10^5 \leq Re \leq 8.56 \times 10^5$ based on $D = 0.025$ m (for the wheel), and $2.05 \times 10^6 \leq Re \leq 5.14 \times 10^6$ based on $D = 0.150$ m (for the cube, see Figure 2). The Reynolds number is given by:

$$Re = \frac{U_\infty D}{\nu} \qquad (6)$$

where $U_\infty$ is the mean flow velocity and $\nu = 1.46 \times 10^{-5}$ m²/s is the kinematic viscosity.

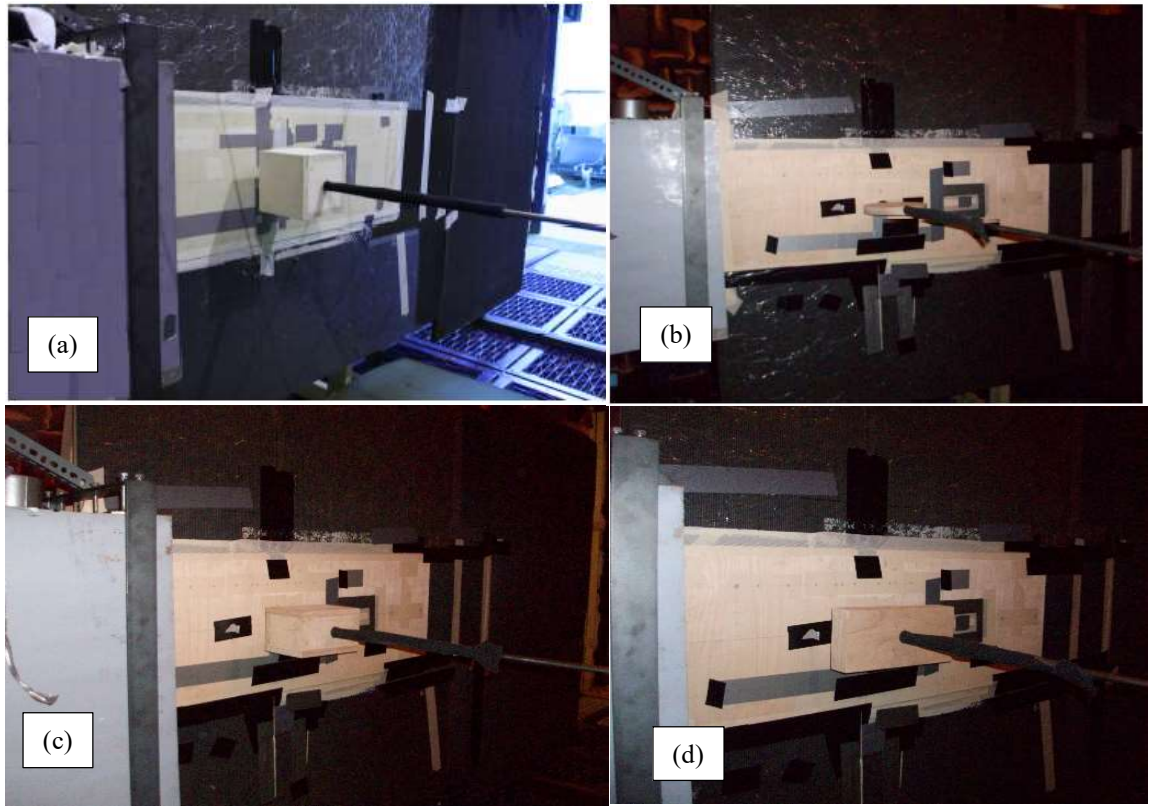

**Fig. 3.** Front view of the experimental set-up used for the noise measurements using simple shapes. (a) Set-up using a cube; (b) set-up using a disc; (c) set-up using a rectangular cuboid; (d) set-up using a long rectangular cuboid. Flow is from left to right.

The orientation of the simple shapes with respect of the incoming flow and the location of the microphone were chosen to be equivalent to the situation during a train pass-by; the exception to this was the rectangular cuboid which was installed with its larger faces parallel to the baffle. This is illustrated in Figure 4 which shows a sketch of a train body and bogie, indicating the location of a microphone at the side of the track during a train pass-by. The orientation with respect to the microphone of the objects used to represent the wheel (disc), the motor (rectangular cuboid) and the gearbox (long rectangular cuboid) is also shown.

Figure 5 shows the noise radiated by the circular strut used to hold the objects, for a flow speed of 31.5 m/s, with and without the addition of the foam. The results show that wrapping porous foam around the strut has suppressed the peak in the noise spectrum due to vortex shedding, in this case at 500 Hz. Although the foam is very efficient to suppress the vortex shedding noise (and hence reduce the background noise for most frequencies) its effect on the flow around the simple shapes was not accounted for. This configuration is considered as the background noise in the experiments.

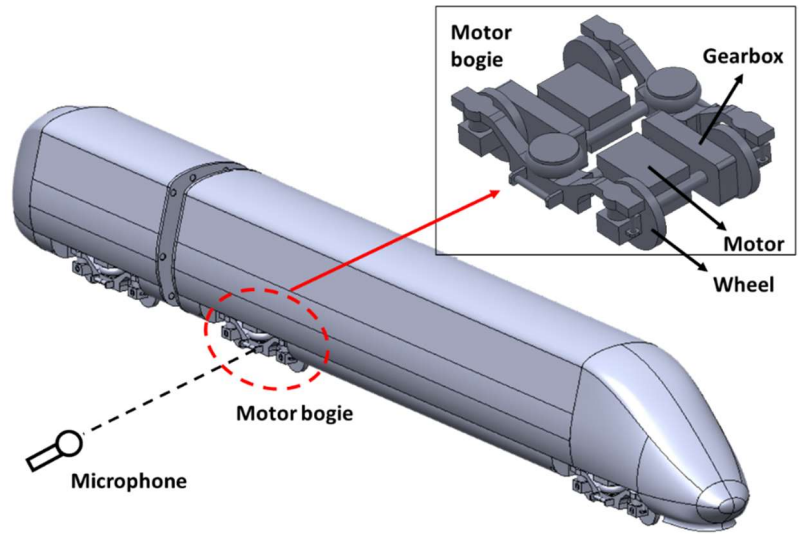

**Fig. 4.** Sketch showing the microphone position during pass-by noise measurements. The relative orientation of the simplified bogie components representing the train wheel (disc), motor (rectangular cuboid) and gearbox (long rectangular cuboid) is also shown.

Figure 6(a) compares the one-third octave band noise spectra measured with the supporting strut wrapped with porous foam and attached to the baffle and when the cube is also included. The signal-to-noise ratio (SNR), defined as the difference between the noise of test configuration being measured and the background noise, is in this case between 3.0 and 10.5 dB in the whole frequency range, except for the frequency band of 10 kHz, with a SNR of 2.5 dB. A SNR between 3.0 and 10.0 dB is sufficient to allow the contribution of the cube to be identified but requires a correction to be applied to allow for the effect of background noise. Figure 6(b) shows the level difference SNR of different simple shapes. For the rectangular cuboid, this is high for frequencies between 315 Hz and 2 kHz, decreasing at lower and higher frequencies, and is lower than 3 dB for frequencies below 125 Hz and above 8 kHz. For the cube, the SNR is between 2.5 and 10 dB in the whole frequency range, while for the long rectangular cuboid it is between 3 and 6 dB for frequencies between 250 Hz and 6.3 kHz, and below 3 dB at lower and higher frequencies. Finally, for the disc the SNR is only greater than 3 dB at the vortex shedding peak, and is lower than this for all other frequencies.

Thresholds of 3 and 10 dB are shown in Figure 6(b). A background noise correction was applied to the frequency bands with a SNR lower than 10 dB, while the frequency bands with a SNR less than 3 dB were discarded (a SNR of 3 dB indicates that the background noise and the noise from object have the same magnitude). For the disc, only the peak noise has a SNR above 3 dB, so the measured data are not reliable for determining the broadband noise from the disc. In any case, this component of the noise can be neglected as it will have no influence on the total noise produced by the bogie. The background noise correction consists of the energetic subtraction of the background noise from the noise measured for the different test configurations. The same procedure was applied for all flow speeds.

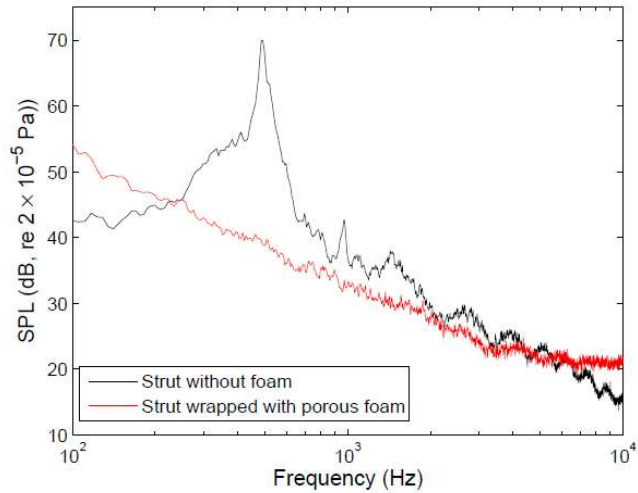

**Fig. 5.** Noise spectra radiated by the circular support strut when it is wrapped with porous foam and without the foam. Flow speed of 31.5 m/s, frequency resolution $\Delta f$ = 5.85 Hz.

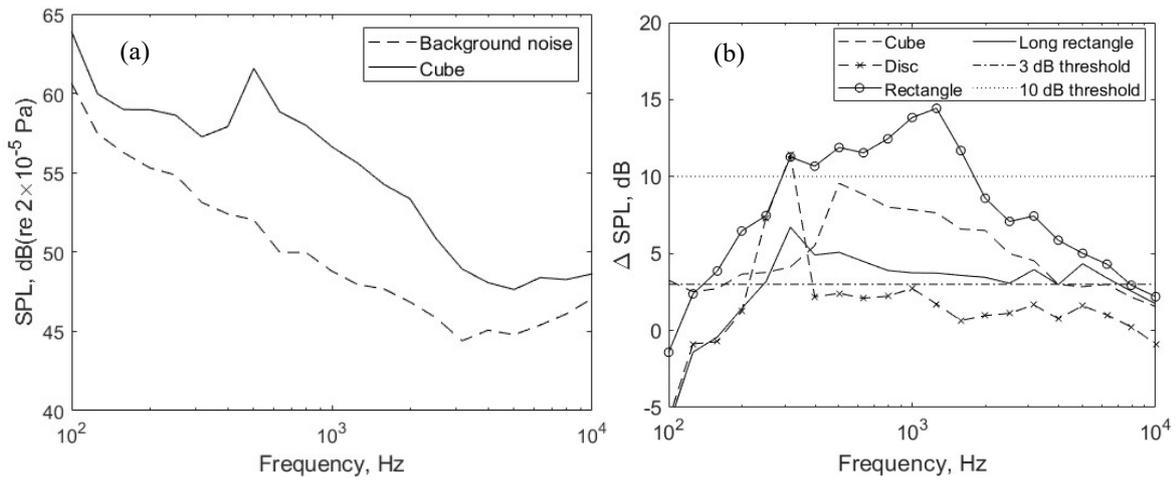

**Fig. 6.** (a) One-third octave noise spectra radiated by the baffle and supporting strut (considered as background noise) alone and with the addition of a single cube. (b) Level difference between the noise spectra measured using different simple shapes and the background noise. Flow speed of 31.5 m/s.

### 3.1.2 Results

Figure 7 shows the measured 1/3 octave band noise spectra for all simple shapes at different flow speeds after applying the background noise correction. For each object, the spectral shape varies with the flow speed: it is relatively flat at low frequencies (except for the disc) and decreases with a nearly constant slope above a specific frequency that is dependent on the flow speed. For the disc, a clear peak can be distinguished, probably due to vortex shedding. Considering the thickness of the disc as the characteristic dimension $D$ (as shown in Figure 2), this peak corresponds to a Strouhal number $St$ = 0.23. The measured values with SNR less than 3 dB (1 dB for the disc) are not included in Figure 7. As the SNR depends on the flow speed and object type, the frequency range covered by the different figure traces is not always the same.

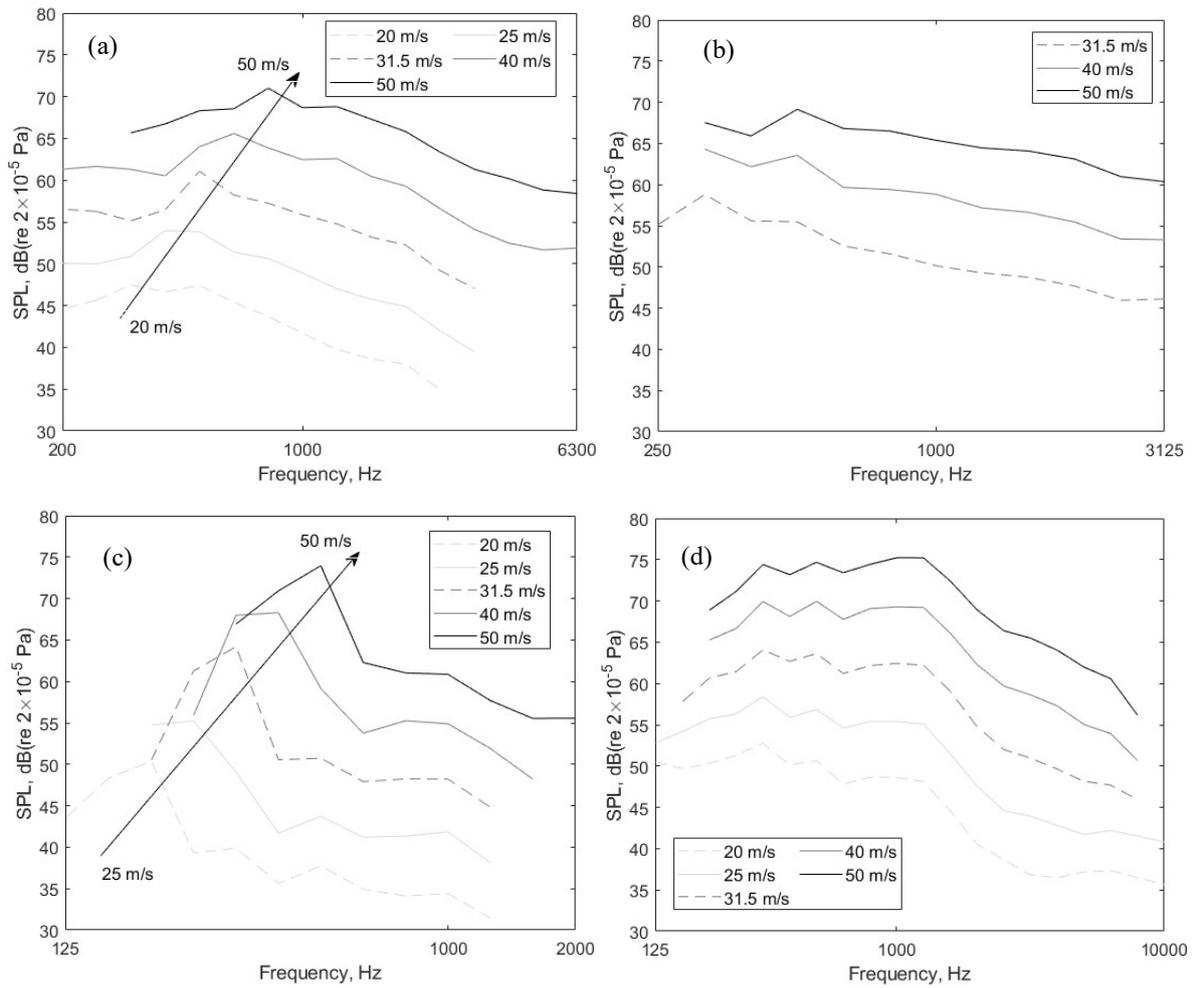

**Fig. 7.** One-third octave band noise spectra measured for different flow speeds. (a) Cube; (b) long rectangular cuboid; (c) disc; (d) rectangular cuboid.

Figure 8 shows the speed dependence of the overall sound pressure level (OSPL) for the various simple shapes. As described above, the noise spectra have first been corrected for the background noise. In the case of the long rectangular cuboid only flow speeds of 31.5, 40 and 50 m/s were used, as for 20 and 25 m/s a significant number of frequency bands had a SNR below 3 dB. The OSPL is calculated by summing over those frequency bands that are retained. The speed exponent $\alpha_i$ is obtained from fitting the OSPL measured for five different flow speeds: 20, 25, 31.5, 40 and 50 m/s against the logarithm of the speed. In each case this value is close to 6.0.

In Figure 9, the measured noise spectra for each flow speed are collapsed in amplitude by subtracting a factor $10\log_{10}(U^{\alpha_i})$ and by plotting the results against Strouhal number $St = fD/U_\infty$. The values used for the exponents $\alpha_i$ are those from Figure 7. The reference spectrum is obtained by averaging the collapsed spectra. This will be used to determine the empirical constants of the prediction model.

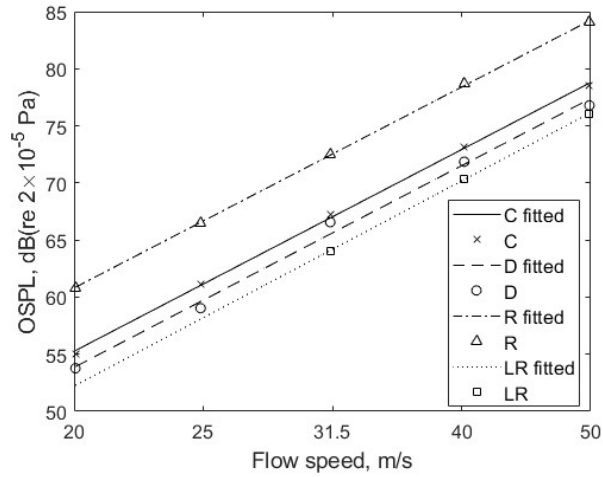

**Fig. 8.** Speed dependence of the measured noise for simple shapes: cube (C), disc (D), rectangular cuboid (R) and long rectangular cuboid (LR). A speed exponent, $\alpha_i$, of 5.9 was obtained for the cube, 6.0 for the long rectangular cuboid, 5.9 for the disc and 5.9 for the rectangular cuboid.

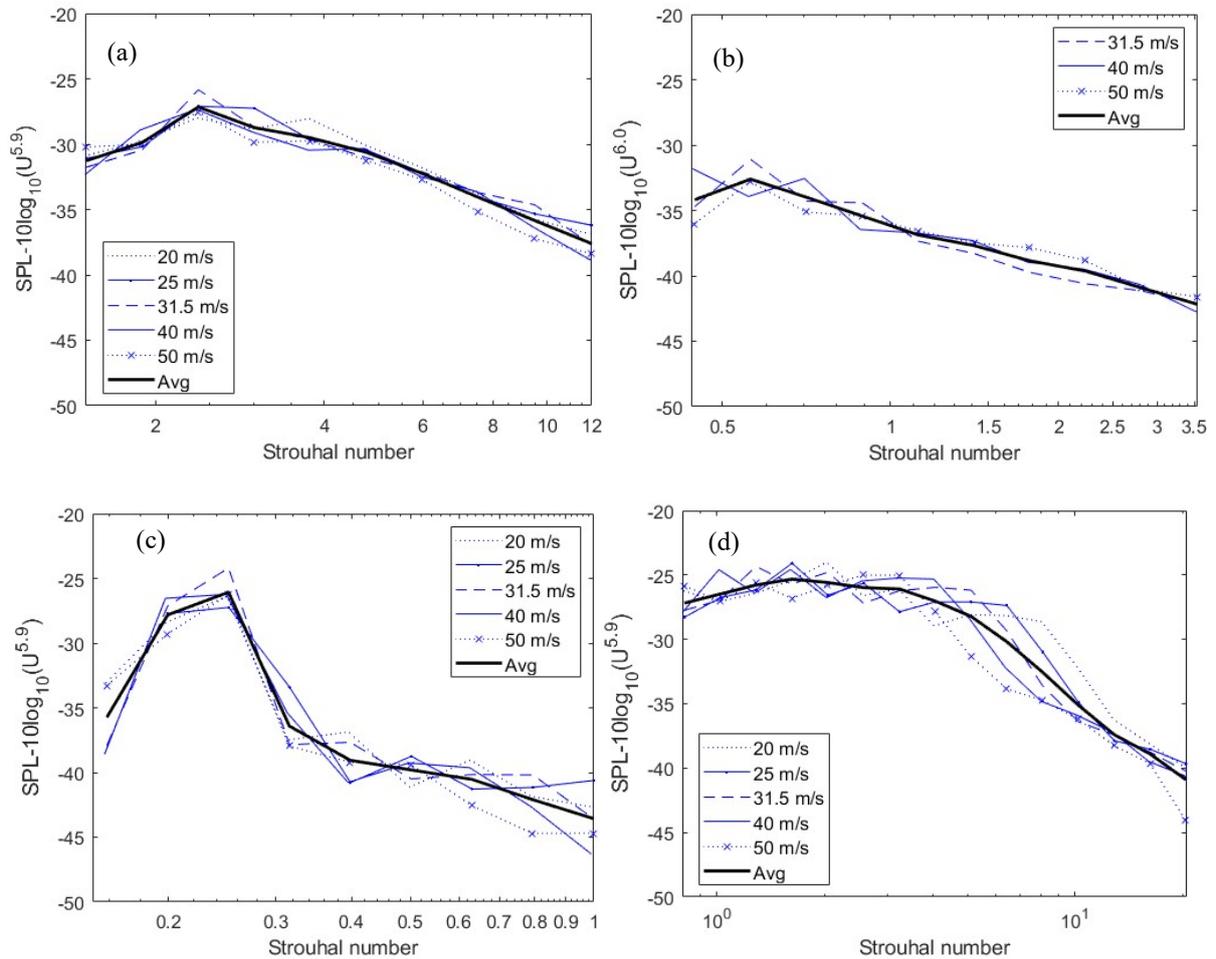

**Fig. 9.** Collapsed noise spectra plotted against Strouhal number. (a) Cube ($U^{5.9}$); (b) long rectangular cuboid ($U^{6.0}$); (c) disc ($U^{5.9}$); (d) rectangular cuboid ($U^{5.9}$).

## 3.2 Adjustment of the component-based prediction model

The value of the amplitude factor, $\eta_i$, and the empirical constants required for the normalized spectrum, $F_i(f)$, are chosen to fit the models in Eqs (2-5) to the reference spectra shown in Figure 9. Tables 1 and 2 show the values of the empirical constants obtained from the fitting. The peak Strouhal number $St_0$ is obtained from the frequency at which the peak of the reference spectrum is located, even where no clear vortex shedding peak is seen. The amplitude factor $\eta_i$ and the empirical constants of the normalized spectrum function are obtained independently for the peak and broadband noise spectra. In addition, the amplitude factor is obtained for an incident flow speed $U_i$ = 31.5 m/s, using the speed exponents $\alpha_i$ shown in Table 1. The specific speed exponent for each of the simple shapes is also introduced in Eq. (3).

The fitting is performed first for the broadband noise spectrum. The amplitude factor $\eta_b$ and the exponents $n_1$ and $n_2$ are adjusted to fit the peak amplitude and width of the predicted spectrum to the broadband part of the reference spectrum. Subsequently, the parameters for the peak noise spectra were chosen: the amplitude factor $\eta_p$ is first chosen to fit the value of the predicted peak to that of the reference spectrum, and subsequently the empirical constant $a_{p2}$ is chosen to obtain an appropriate bandwidth. This procedure was made by eye.

**Table 1.** Values of the empirical constants obtained from the fitting of the predictions with the reference spectrum for each of the simple shapes. Peak noise.

| Parameter | Disc | Long rect. | Rectangle | Cube |
|---|---|---|---|---|
| $St_0$ | 0.23 | 0.56 | * | 2.40 |
| $\alpha$ | 5.9 | 6.0 | * | 5.9 |
| $a_{p2}$ | 0.041 | 0.725 | * | 0.150 |
| $\eta_p$ | $0.24 \times 10^{-5}$ | $2.6 \times 10^{-5}$ | * | $0.02 \times 10^{-5}$ |
| $S_i$ (m²) | 0.072 | 0.113 | 0.115 | 0.135 |

**Table 2.** Values of the empirical constants obtained from the fitting of the predictions with the reference spectrum for each of the simple shapes. Broadband noise.

| Parameter | Disc | Long rect. | Rectangle | Cube |
|---|---|---|---|---|
| $n_1$ | 1.9 | 2.5 | 2.25 | 2.0 |
| $n_2$ | 1.8 | 2.0 | 1.75 | 2.0 |
| $St_0$ | 0.23 | 0.56 | 2.25 | 2.40 |
| $\eta_b$ | $0.09 \times 10^{-5}$ | $0.16 \times 10^{-5}$ | $0.96 \times 10^{-5}$ | $0.20 \times 10^{-5}$ |
| $\alpha$ | 5.9 | 6.0 | 5.9 | 5.9 |
| $S_i$ (m²) | 0.072 | 0.113 | 0.115 | 0.135 |

Figure 10 shows a comparison between predicted and measured spectra for the different simple shapes at a flow speed of 31.5 m/s. There is good agreement between the measurements corrected for background noise and the fitted curves. For the rectangular cuboid, the peak noise component can be neglected as the broadband noise spectrum matches the measurement. The differences in overall level between the predictions and measurements are mostly less than 1 dB for all objects and all flow speeds.

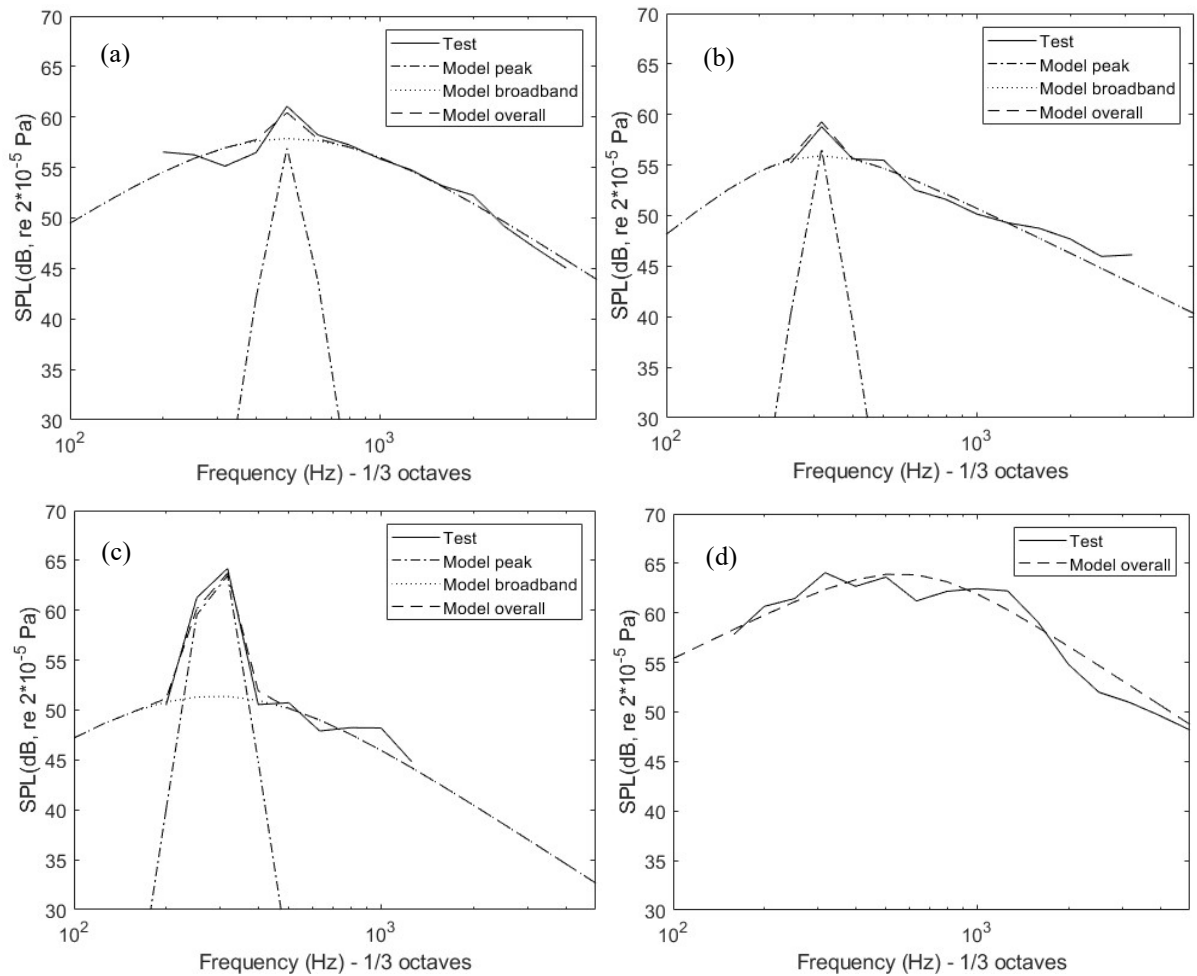

**Fig. 10.** Comparison between experiments and predictions for different simple shapes at flow speed 31.5 m/s. (a) Cube; (b) long rectangular cuboid; (c) disc; (d) rectangular cuboid.

Only a single microphone was used in the measurements, which was placed along the line perpendicular to the flow direction from the centre of the simple shapes, as shown in Figure 1. Measurements of the radiated noise from different angles would be required to estimate the directivity function $D_{rad_i}(\psi,\phi)$, so this is beyond the scope of this work and could be addressed in future experiments.

## 4 Experiments and predictions for 1:10 scale bogie mock-up

A second set of experiments was carried out with a 1:10 scale simplified bogie with removable components made of high-density stiff foam. Some components were added or removed to assess their effect on the noise. This also allowed a trailer and a motor version of the simplified bogie to be tested. Figure 11 shows a sketch of the motor and trailer bogie configurations, in which the main components are identified. The use of a scale factor of 1:10 implies that the frequency of the noise spectra is 10 times higher than those for a full-scale bogie, while the sound pressure amplitudes are reduced. For comparison with full-scale quantities, Eq.(1) can be used to adjust the amplitude.

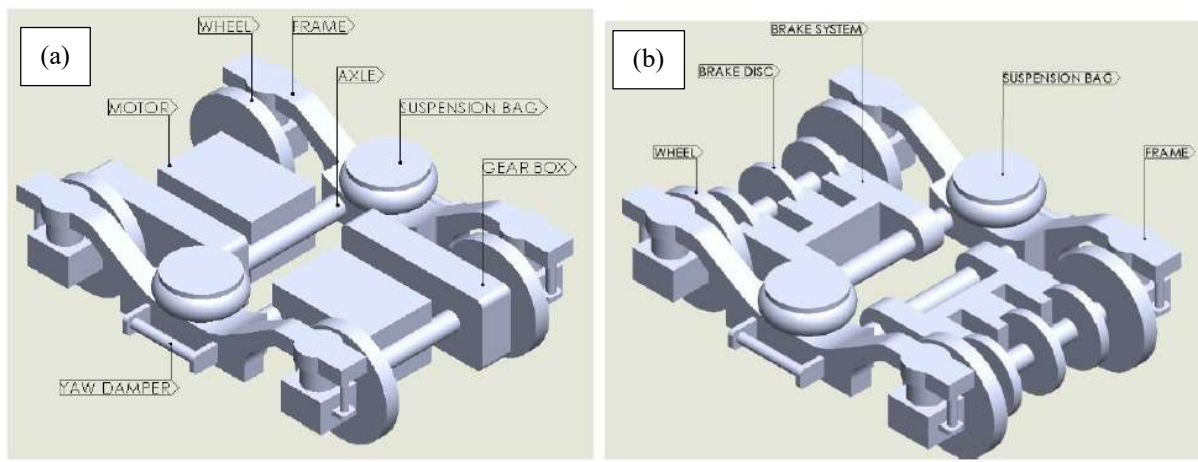

**Fig. 11.** Sketch of the bogie mock-ups used. The geometry of the bogie components is approximated by simple objects. (a) Motor bogie; (b) trailer bogie.

### 4.1 Experimental set-up

The bogie was installed on the same plate in the baffle, see Figure 1. It was installed in two configurations, shown in Figure 12: in one it was simply mounted on the flat plate (Figure 12(a)) and in the other it was installed between two curved ramps, used to approximate a simplified bogie cavity (Figure 12(b)). In both cases, the noise from both the motor and trailer bogie configurations was measured.

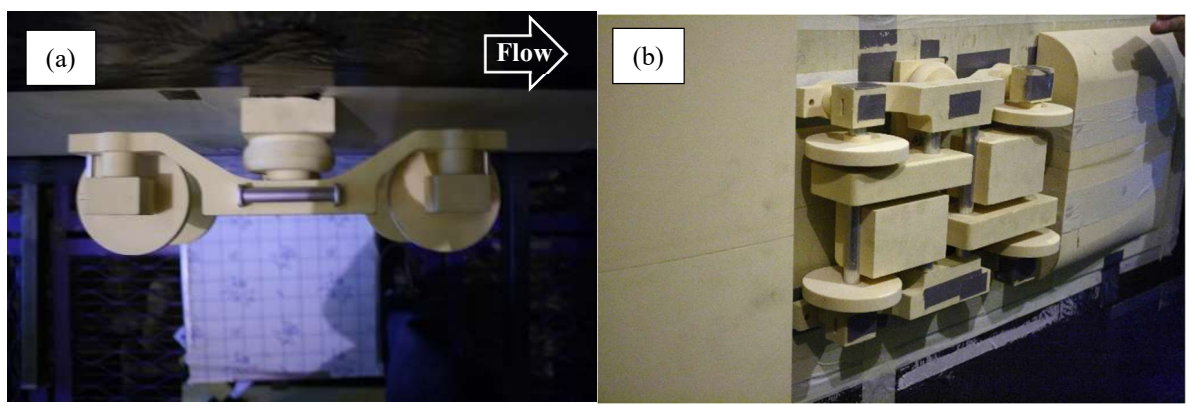

**Fig. 12.** 1:10 scale simplified bogie attached to the baffle. (a) Without ramps; (b) with ramps.

Figure 13 shows the test set-up with the bogie and ramps attached to the baffle. The upstream ramp was wider than the nozzle to avoid separation at the edges and to ensure the flow incident on the bogie mock-up is laminar.

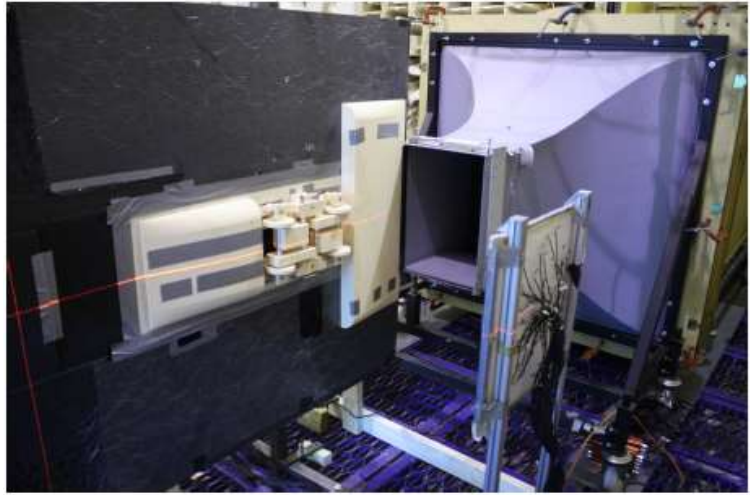

**Fig. 13.** Experimental set-up showing the 1:10 scale simplified bogie inside a simplified bogie cavity.

**4.2. Results**

Figure 14(a) shows examples of the noise spectra measured for a flow speed of 50 m/s. Results are shown for the baffle alone, the baffle with the ramps, the motor bogie on the baffle with ramps, and the trailer bogie on the baffle with ramps. Compared with the result for the baffle alone, the signal-to-noise ratio is greater than 7 dB in the whole frequency range for both the motor and trailer bogie. However, compared with the case of the baffle and ramps, the signal-to-noise ratios are smaller. Figure 14(b) shows the level difference between the noise measured with the motor or trailer bogie installed and for the case with the ramps alone at different flow speeds. In all cases, except at low frequency, the differences are greater than 3 dB. From this it can be inferred that the noise from the bogie is higher than that from the ramps.

Figure 15 shows the noise measured for both the motor and trailer bogie configurations without the ramps, for flow speeds of 25, 40 and 50 m/s. Both axles are present in each case, but the components attached to them (motor and gearbox or brake discs) are removed from each axle in turn. Results are compared for configurations in which the bogie components are installed on both axles, only on the front axle (bare rear axle) and only on the rear axle (bare front axle). When the front axle is exposed, two prominent peaks appear, at a frequency that increases with flow speed, probably due to vortex shedding from the axle. These peaks disappear when the components are installed on the front axle. No obvious peaks are found when the rear axle is exposed.

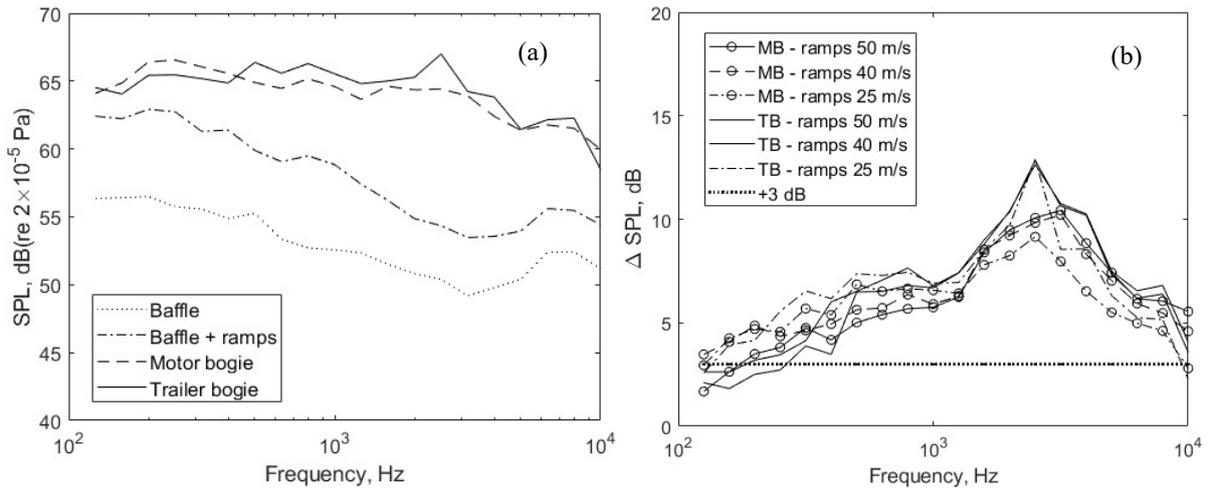

**Fig. 14.** (a) Noise spectra measured at 50 m/s for: baffle alone (background noise); baffle and ramps; baffle, ramps and motor bogie; and baffle, ramps and trailer bogie. (b) Difference between the noise spectra measured with ramps alone and with ramps and the motor or trailer bogie at different flow speeds.

Figure 16 shows the corresponding results when the two ramps are installed, meaning that the bogie is located in a simplified cavity. Compared with the results in Figure 15, the sound levels are generally about 5 dB lower across most of the frequency range. No strong tonal peaks are apparent when the front axle is exposed. For frequencies below 1250 Hz, the noise when the components are installed on both axles is slightly lower than when they are installed only on one axle. Above this frequency, the noise with the front axle exposed is lower than the other two configurations, these giving very similar results.

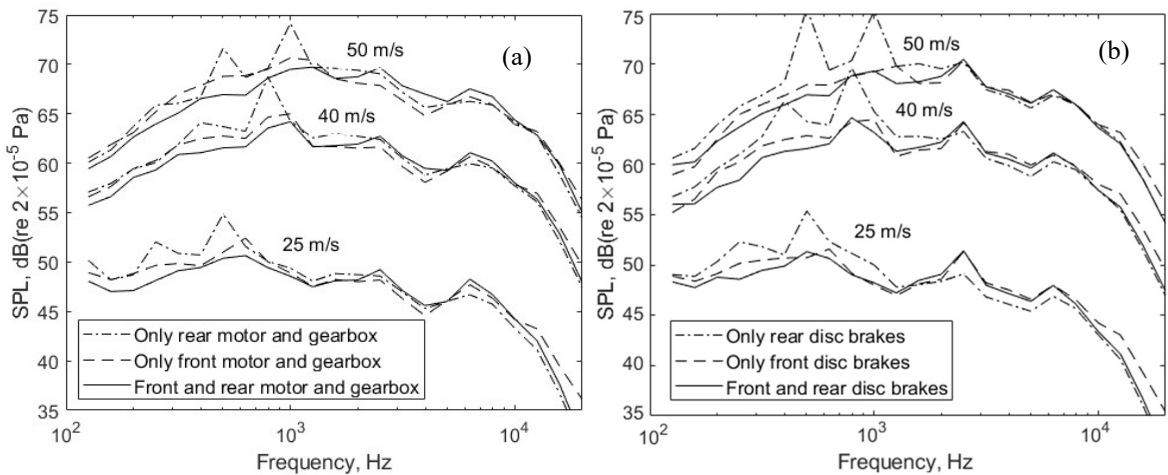

**Fig. 15.** Noise measured for the bogie with the components installed on both axles, just on the front axle (bare rear axle) and just on the rear axle (bare front axle), for flow speeds of 25, 40 and 50 m/s and no ramps. (a) Motor bogie; (b) trailer bogie.

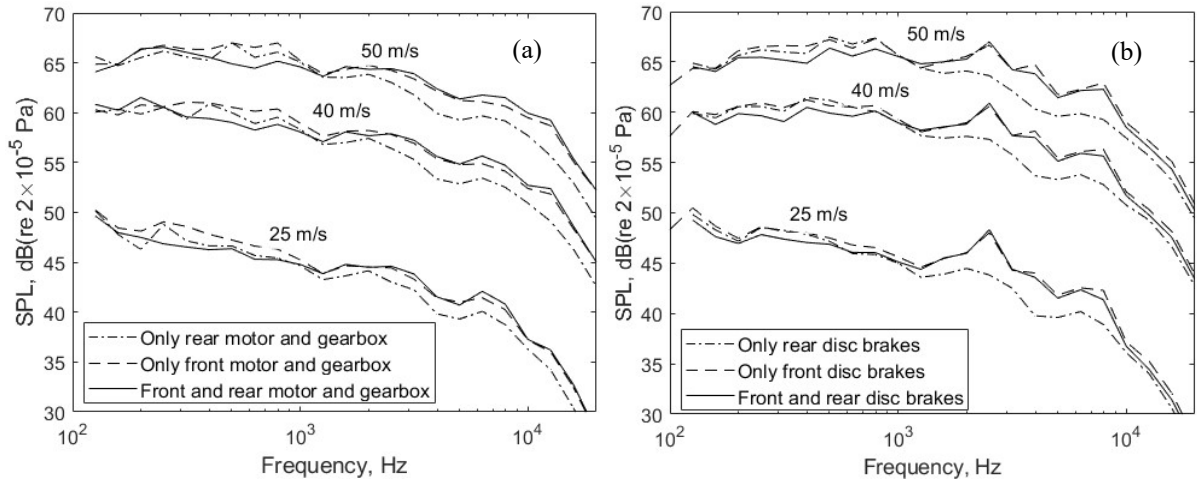

**Fig. 16**. Noise measured for the bogie with the components installed on both axles, only on the front axle or only on the rear axle, for flow speeds of 25, 40 and 50 m/s and with ramps. (a) Motor bogie; (b) trailer bogie.

## 5   Comparison between measurements and predictions

In this section the measurement results for the 1:10 scale bogie mock-up are compared with predictions in which the bogie components are represented by the simple shapes according to their geometry. Cases with and without the upstream and downstream ramps are considered. In the case with the ramps, two effects have to be considered: the self-noise from the ramps themselves and the flow screening effect due to the upstream ramp. Flow speeds of 25, 40 and 50 m/s are considered. As in the previous section, the frequency scale shown in all the figures in this section is as it was measured, which is 10 times higher than at full scale.

For simplicity, only the front components of the bogie are included in the prediction model, those components being highlighted in Figure 17. This simplification is based on the observation that the noise from the complete bogie and the bogie with exposed rear axle is nearly the same, as shown in Figure 15 for the bogie alone and Figure 16 for the bogie with ramps. This observation implies that the upstream components are shielding the flow from the downstream components, reducing the incident flow speed, especially for the case without ramps. All the comparisons are made for the microphone position perpendicular to the flow direction from the bogie centreline so the effects of the directivity and convective amplification can be neglected.

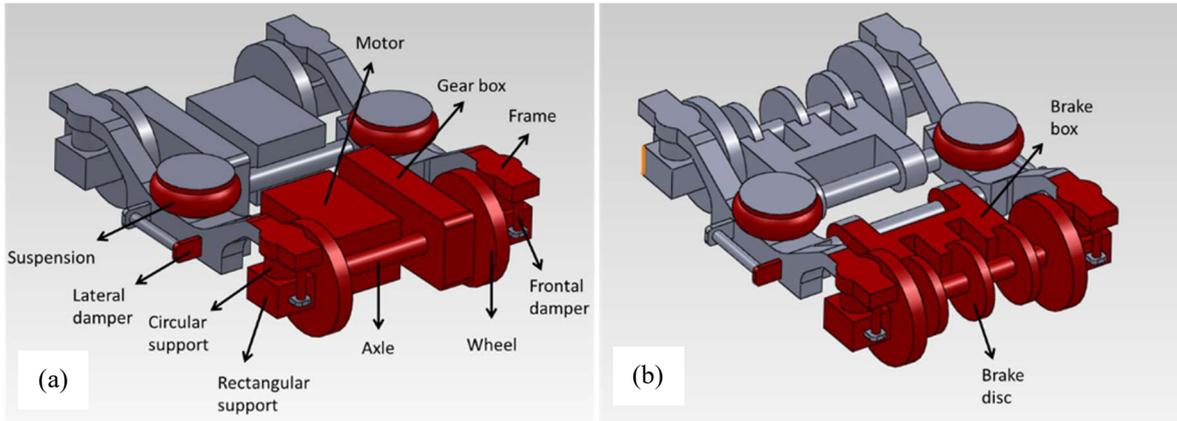

**Fig. 17**. Sketch of the bogie mock-ups. The components included in the prediction model are highlighted. (a) Motor bogie; (b) trailer bogie.

### 5.1 Bogie alone

In the case of the bogie alone, with no ramps, the front components are exposed to the incoming flow. The incident flow speed for all these components is assumed to be equal to the mainstream flow speed $U_\infty$ and the incoming turbulence $I_u$ is neglected, to approximate the conditions during the experiments. The dimensions of each component, and the objects used to approximate their geometry, are given in Table A.1 in the Appendix.

Figure 18 shows a comparison between the predicted and measured noise spectra for the motor and trailer bogie configurations for a flow speed of 50 m/s. For the motor bogie (Figure 18(a)), the difference between predictions and measurements is 0.6 dB in terms of OSPL, while for flow speeds of 25 and 40 m/s (not shown) differences of 0.8 and 0.5 dB were obtained. The predicted spectral shape is also in good agreement with the measurements apart from a prominent peak at 5 kHz in the measurements that is not predicted by the model. The frequency of this peak is found to be independent of the flow speed, so it is probably not related to the noise produced by any of the components of the bogie. According to the predictions, the noisiest component for the motor bogie is the motor, which has the largest frontal area. The axle box also contributes to the overall noise, mainly at high frequencies, and the noise from the gearbox is significant at mid frequencies. Several other components are important at low frequencies including the suspension, gearbox and axle.

In the case of the trailer bogie, a good agreement is again found between the predicted and measured OSPL, with a maximum difference of 0.8 dB for a flow speed of 50 m/s. However, the differences in the spectral shape are more significant, as shown in Figure 16(b). In this case the noise source with the highest contribution at low frequencies is the brake system, while for mid and high frequencies the contribution from the axle boxes is again significant.

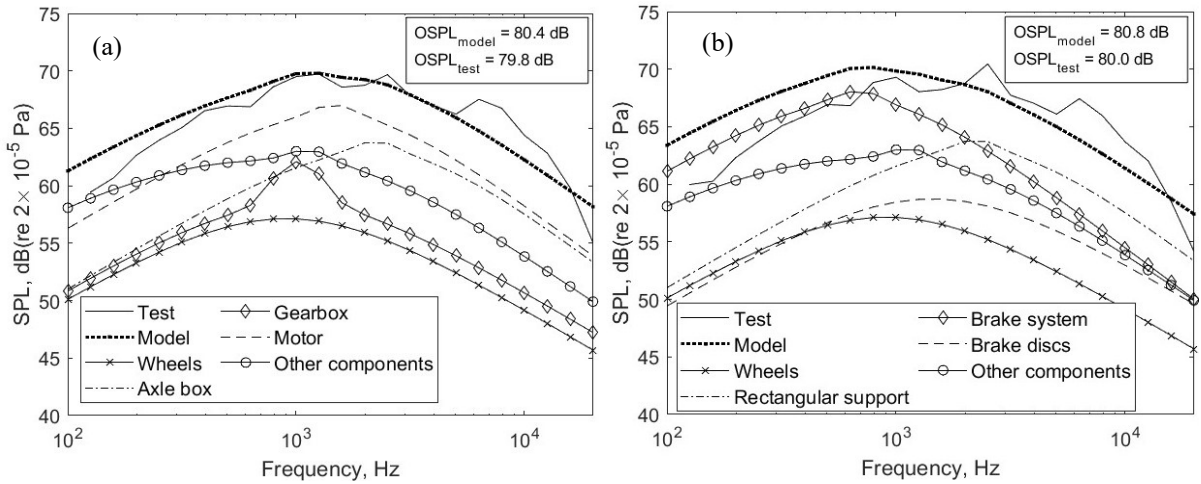

**Fig. 18**. Comparison between the measured and predicted sound pressures at 50 m/s when the ramps are not attached. (a) Motor bogie; (b) trailer bogie. Frequency range as measured.

Figure 19 shows the level differences between the predicted and measured noise spectra. Results are shown for the motor and trailer bogie without ramps at flow speeds of 25, 40 and 50 m/s. For frequencies below 2 kHz the noise is slightly overpredicted, while for frequencies above this the noise is underpredicted. The noise difference is between -4 and +5 dB in the whole frequency range. The differences in the OSPL are between 0 and +1 dB for all cases. The consistency of these results indicates that the speed dependence used in the predictions also fits the experimental results.

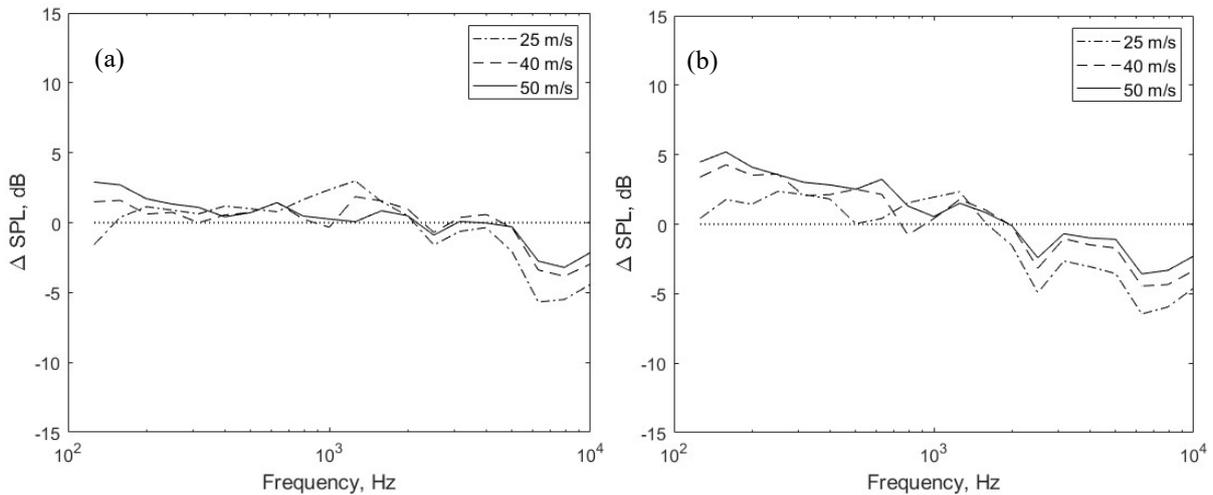

**Fig. 19.** Differences between the predicted and measured noise spectra for the bogie without ramps at flow speeds of 25, 40 and 50 m/s. (a) Motor bogie; (b) trailer bogie.

### 5.2 Bogie in cavity

When the bogie is installed in the cavity between the ramps, the size and geometry of the components included in the prediction model is based on the parts of the motor and trailer bogie configurations that are exposed to the incident flow. Figure 20 shows the relative position of the bogie components and the ramps, indicating the parts of the components exposed to the incoming air flow and the parts shielded by the ramps. Only the parts that are exposed to the flow are included in the model.

Due to the inclusion of the ramps, it may be appropriate to use different normalised spectra for some of the components. However, without further evidence no adjustments were made. The size of the components and the objects chosen to represent them are given in Table A.2 in the Appendix. For simplicity the incident flow speed for all these components is assumed to be the mainstream flow speed $U_\infty$ and the incoming turbulence $I_u$ is neglected. Further work is required to generalise the approach for modelling the components based on their relative position in the bogie cavity so it can be applied to different bogie geometry.

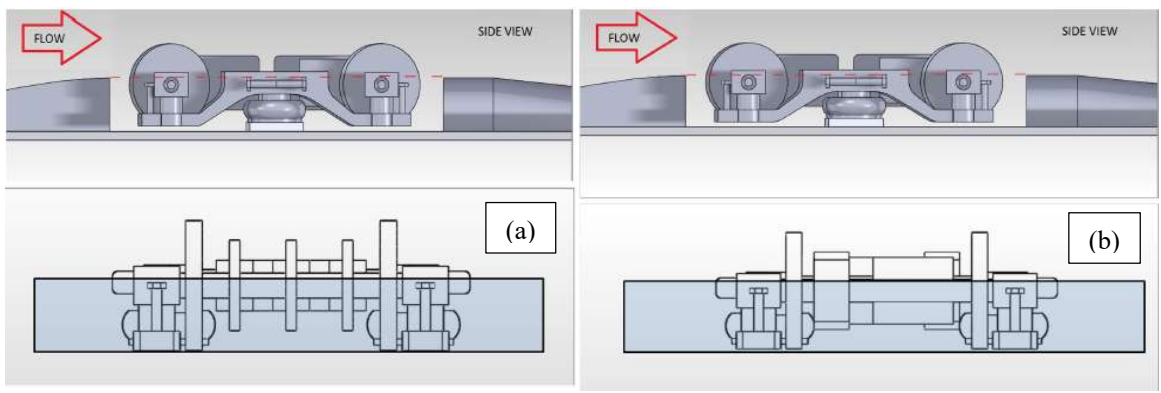

**Fig. 20.** Sketch showing the position of the different components of the bogie with respect to the upstream ramp. The bogie model has a maximum length of 344.0 mm and maximum width of 277.9 mm. The upstream ramp has a length of 300.0 mm, a width of 700.0 mm and a height of 60.0 mm. The components of the bogie that are not shielded by the upstream ramp can be distinguished from those which are shielded. (a) Trailer bogie; (b) motor bogie.

The results obtained for the motor bogie configuration are shown in Figure 21(a). The predicted noise is slightly higher than the measurements in the mid frequencies. The motor remains the component with the highest contribution to the overall noise. The hump at 5 kHz is not predicted by the model, similar to the result found when the ramps were not included. However, unlike the case without ramps, the predicted noise at low frequencies is significantly lower than the measurements. The measured noise for the case without the bogie (empty cavity) is also shown and is significant for frequencies below 315 Hz. As the noise from the cavity is not included in the model, this seems to be the main reason for the disagreement between prediction and measurement at low frequencies.

The same trend is found for the trailer bogie, Figure 21(b). The noise at low frequencies is under-predicted due the noise produced by the cavity. In the measured data a prominent peak appears at 2.5 kHz, which is not found in the predictions. This frequency remains unchanged with the flow speed and is believed to be unrelated to the bogie components. The main noise source from the bogie at low frequencies is the brake system, the axle boxes also being significant at mid and high frequencies.

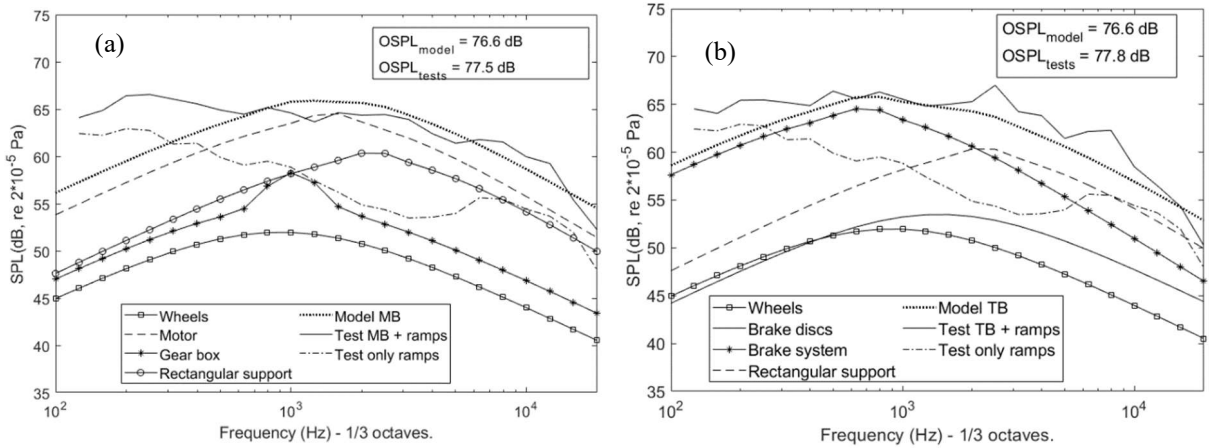

**Fig. 21**. Comparison between measured and predicted sound pressures at 50 m/s for the bogie configurations with ramps. (a) Trailer bogie; (b) motor bogie. Frequency range as measured.

Figure 22 shows the differences between the predicted and measured noise spectra for flow speeds of 25, 40 and 50 m/s. For frequencies above 630 Hz, in the case of the motor bogie, the results show good agreement between predictions and measurements (differences less than 2.5 dB), while for the trailer bogie the noise is underpredicted, with a maximum difference of 4 dB. For frequencies below 630 Hz the noise is underpredicted for both bogie configurations, the discrepancy increasing as the frequency decreases. Nevertheless, the differences in the OSPL are between -1 and 0 dB for all cases.

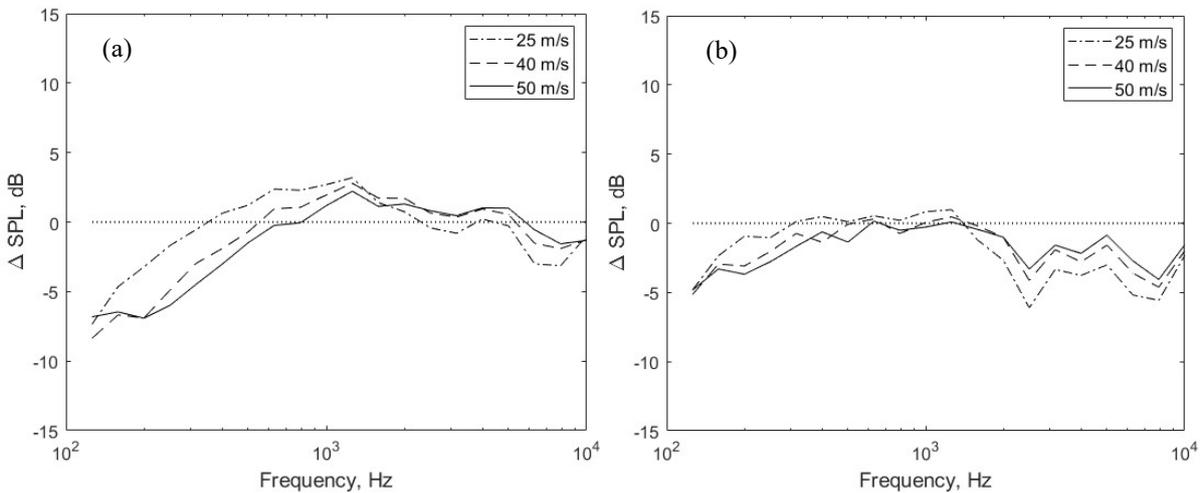

**Fig. 22**. Differences between the predicted and measured noise spectra for the motor and trailer bogie with ramps at. flow speeds of 25, 40 and 50 m/s. (a) Motor bogie; (b) trailer bogie.

5.3 Discussion

From these results it can be concluded that the model is working well to predict the noise of the bogie alone, that is, directly exposed to the incoming flow. However, for the case of the bogie inside a cavity the model underpredicts the noise in the low frequency region. However, this region corresponds to frequencies below around 50 Hz at full scale; for speeds up to 100 m/s it could extend up to 100 Hz. It is believed that this disagreement is due to the noise produced by the cavity itself, which is not predicted by the model (see Fig. 21).

Several recommendations are given that could improve the prediction model accuracy: the geometry of each component in the model should take account of its relative position with respect to the bogie cavity. Where components are partially shielded by the upstream bogie cavity step, the geometry exposed to the flow will differ from that of the whole component. Moreover, the effect of incoming turbulence, as well as the interaction between bogie components should be considered. In addition, for the application of the model to a practical case, omission of the small details could lead to an underprediction of the high-frequency noise.

In addition, further investigations are required to include the effect of the ground in the measurement set-up to obtain flow conditions more representative of those in a real train bogie cavity. The distance between the different simple shapes and the ground and their relative position inside the bogie cavity should also be considered as they can have a non-negligible effect on the radiated noise. The estimation of the directivity function $D_{rad_i}(\psi, \phi)$ of the different simple shapes could also be addressed in future experiments by adding additional microphones at different angles with respect the shape axis perpendicular to the flow direction.

## 6. Conclusions

A semi-empirical component-based model originally developed for the aerodynamic noise from train pantographs, consisting of cylinders, has been extended for application to train bogies consisting of bluff bodies. Anechoic wind tunnel noise measurements were carried out using simple shapes representing bogie components. The empirical constants in the model were adjusted for application to the bogie case by fitting the model to reference spectra obtained after collapsing in amplitude the measured noise spectra for different flow speeds ranging from 20 to 50 m/s.

Predictions from this model have been compared with noise measurements on a 1:10 scale bogie mock-up in motor and trailer bogie configurations. This was installed on a baffle plate on its own and between two ramps representing a simplified bogie cavity. In the case of the bogie alone, predictions showed good agreement with measurements, with overall sound pressure levels within less than 1.2 dB of the measurements for all the flow speeds tested and with spectral differences between -5 and 5 dB in the whole frequency range. When the bogie was installed between the ramps, the OSPL differences between predictions and measurements were less than 1.0 dB for all flow speeds but the spectral differences increased to 7 dB at low frequencies, showing that the low-frequency noise is underpredicted. This low-frequency noise is attributed to the cavity and will be concentrated below 100 Hz on a full-scale train.

The results obtained show the feasibility of the application of the semi-empirical component-based model to the bogie case. The prediction can be improved if the definition of the objects used in the model is extended to consider the screening effect of the cavity. Account should also be taken of the effect of the incoming turbulence, variations in the incident flow speed, interactions between components and the inclusion of small geometrical details.

**Funding**: This work was supported by the Minsterio de Ciencia e Innovación, the Agencia Estatal de Investigación and the FEDER [grant number PID2021-122237OA-I00]. In addition, J. García would like to thank the Programa de Excelencia para el Profesorado Universitario de la Comunidad de Madrid for their financial support.

# Appendix A

The dimensions of each component and the simple shape that approximates its geometry for the bogie alone are given in Table A.1. MB stands for motor bogie and TB stands for trailer bogie.

**Table A.1.** Main dimensions and type of normalized spectrum chosen for each of the 1:10 scale bogie components included in the prediction model. Dimensions in mm. Bogie without ramps. The inclusion of the component in the trailer bogie (TB), motor bogie (MB) or both of them, is shown using the marker X. For the length of the wheel axle the value between brackets corresponds to the case of the trailer bogie where the axle is considered as two independent shorter cylinders to account for the effect of the brake discs.

|  | D (mm) | B (mm) | L (mm) | Norm. spectrum | MB | TB |
|---|---|---|---|---|---|---|
| **Wheel shaft** | 17.5 | - | 115.0 (2 x 45) | Circular cylinder | X | X |
| **Circular support** | 27.0 | - | - | Cube | X | X |
| **Axle box** | 35.0 | 40.0 | 27.0 | Rectangle | X | X |
| **Lateral frames** | 25.0 | 14.0 | 64.0 | Long rectangle | X | X |
| **Wheels** | 12.5 | 92.0 | - | Disc | X | X |
| **Motor** | 53.0 | 64.0 | 94.0 | Rectangular | X | - |
| **Gearbox** | 28.0 | 140.0 | 61.0 | Long rectangular | X | - |
| **Brake system** | 114.0 | 96.0 | 36.0 | Rectangular | - | X |
| **Brake discs** | 8.0 | 64.0 | - | Discs | - | X |
| **Lateral damper** | 25.0 | 16.0 | 4.0 | Long rectangular | X | X |
| **Vertical damper** | 9.0 | - | 62.0 | Circular cylinder | X | X |
| **Suspension** | 70.0 | 62.0 | - | Disc | X | X |

**Table A.2.** Main dimensions and type of normalized spectrum chosen for each of the 1:10 scale bogie components included in the prediction model. Dimensions in mm. Bogie with ramps. The dimension $B$ is expressed using fractions to indicate the part of the component that is not shielded by the upstream ramp in relation with the total size of the component.

|  | D (mm) | B (mm) | L (mm) | Norm. spectrum |
|---|---|---|---|---|
| **Axle box** | 35.0 | 40.0/4 | 27.0 | Rectangle |
| **Wheels** | 12.5 | 92.0/2 | - | Disc |
| **Motor** | 53.0 | 64.0/3 | 94.0 | Rectangular |
| **Gearbox** | 28.0 | 140.0/3 | 61.0 | Long rectangular |
| **Brake system** | 114.0 | 96.0/3 | 36.0 | Rectangular |
| **Brake discs** | 8.0 | 64.0/2 | - | Discs |